\documentclass[11pt,letterpaper]{article}
\usepackage[margin=1.08in]{geometry}
\usepackage{amsmath}
\usepackage{setspace}
\usepackage{authblk}
\begin{document}
\renewcommand{\d}{\mathrm{d}}
\title{The Brute-Force Search for Planet Nine}
\author[1]{Scott Lawrence\footnote{srl@umd.edu}}
\author[2]{Zeeve Rogoszinski\footnote{zero@umd.edu}}
%\email{srl@umd.edu}
\affil[1]{Department of Physics, University of Maryland, College Park, Maryland 20742, USA}
\affil[2]{Department of Astronomy, University of Maryland, College Park, Maryland 20742, USA}
\maketitle
\begin{abstract}
A recent proposal~\cite{witten1} for the detection of a hypothetical gravitating body $\sim 500\;\mathrm{AU}$ from the Sun (termed Planet 9) calls for a fleet of near-relativistic spacecraft, equipped with high-precision clocks, to be sent to a region where the object is suspected to be. We show that the technological constraints of such a mission can be relaxed somewhat, while improving the sensitivity: high-precision clocks can be avoided when the transverse displacement induced by Planet 9 is measurable with Earth-based, or near-Earth, telescopes. Furthermore, we note that in the absence of Planet 9, these spacecraft still yield useful data by mapping gravitational perturbations in the outer parts of the solar system.
\end{abstract}

Gravitational anomalies have a long history of being used in the discovery of unseen objects in the solar system. Most recently, evidence of apsidal clustering of extreme trans-Neptunian objects (eTNOs) has suggested the existence of an unseen body, termed Planet 9~\cite{Batygin_2016,Batygin_2019}. The mass of this body is estimated to be $5-10 M_\oplus$, and the distance from the Sun on the order of $500\;\mathrm{AU}$. It has additionally been suggested that this body may be a primordial black hole or another exotic object, and therefore exceptionally worthy of direct study~\cite{scholtz2019planet}. Unfortunately, direct searches for Planet 9 have so far yielded nothing. If the body is in fact a compact exotic object, or if observed eTNO clustering is instead a byproduct of another mechanism such as inclination instability~\cite{2016MNRAS.457L..89M,2020arXiv200401198Z}, then it is likely that direct searches will continue to be fruitless.

Indirect searches for the body, which proceed by detecting its gravitational influence, are somewhat more promising. For example, the Cassini spacecraft's measurement of Saturn's orbit places some constraints on the mass and location of Planet 9~\cite{Fienga_2016,2016AJ....152...94H}. Hints of a possible Planet 9 were seen in these measurements, but these perturbations to Saturn's ephemeris can also be accounted for by accelerations from the total mass contained in the classical Kuiper Belt ($M_{\mathrm{belt}}\approx0.02M_{\oplus}$)~\cite{2018CeMDA.130...57P}.

Such experiments depend on an existing visible object being measurably affected by the gravitational influence of Planet 9. A more aggressive search without this constraint was recently suggested~\cite{witten1}, in which a fleet of near-relativistic spacecraft are launched into the region suspected to contain Planet 9. Passing through the gravitational field of the body would cause one to speed up temporarily, revealing the location of the object.
This proposal depends critically on several advances in technology. First, in order for the mission to be completed in a reasonable timespan, the probes are constrained to be travelling at a speed of order $0.001\,c$ or greater, more than an order of magnitude over the speed achieved by the New Horizons spacecraft. Current proposals in this vein (eg. ``Breakthrough Starshot''~\cite{starshot}) require the spacecraft to have total mass of only a few grams. Second, the detection of the temporary speed-up of the probe requires the probe to carry a high-precision clock, accurate to about one part in $10^{12}$ ($10^{-5}$ seconds over about one year). No such clock suitable for an ultralight spacecraft currently exists, and indeed this is considered in~\cite{witten1} to be a primary obstacle.

Here we discuss a similar method, which again depends on a fleet of near-relativistic probes, but removes the need for any high-precision clock. Instead, the position of the probes in the sky is measured to high precision via Very Long Baseline Interferometry (VLBI), which has previously been used in tracking near-Earth satellites~\cite{lanyi2007angular,hanada2008vlbi}. The entire task of the probe is to be visible via radio waves, serving as a test particle for an experiment observed from Earth. Only the transverse displacement can be measured in this way, but the extent of transverse displacement is relatively large. This method has in common with a previous proposal~\cite{loeb} the use of interferometry to precisely measure acceleration, but here the interferometer remains on or near Earth, reducing the complexity and weight of the spacecraft.

As in~\cite{witten1}, we consider the trajectory of a spacecraft incident on Planet 9, evaluated to first order in the perturbation introduced by the gravity of Planet 9 (the impulse approximation). With incident velocity $v^{(0)}$ taken to be along the $x$-axis, and impact parameter $\rho$, the unperturbed trajectory is given by
\begin{equation}\label{eq:unperturbed}
(x^{(0)}(t), y^{(0)}(t),z^{(0)}(t)) = (v^{(0)}_x(t), \rho, 0)
\text.
\end{equation}
In the original proposal, the perturbation to $v_x$ gave rise, after the spacecraft passed Planet 9, to a permanent displacement in the radial ($x$) direction. Here we consider instead the transverse ($y$) direction. In passing a body of mass $M$, the velocity is perturbed according to
\begin{equation}
\frac{\d v_y}{\d t}
\approx
-
\frac
{G M y^{(0)}(t)}
{(x^{(0)}(t)^2 + y^{(0)}(t)^2)^{3/2}}
\text.
\end{equation}
Integrating over all times using the trajectory (\ref{eq:unperturbed}), we see that this perturbation induces a permanent change in the transverse velocity, inversely proportional to the impact parameter and launch speed of the probe:
\begin{equation}
\Delta v_y = \frac{2 G M}{\rho v^{(0)}_x}
\text.
\end{equation}
As a result, a transverse displacement $\Delta y(t) \approx \Delta v_y t$ builds up starting when the spacecraft is in the vicinity of Planet 9. As the spacecraft is, to a good approximation, moving directly away from the earthbound observer, this transverse displacement corresponds to an angular displacement in the sky of $\Delta y / D$, where $D$ is the distance to the spacecraft at the time the displacement is measured. The spacecraft continues to move away from the observer while the displacement builds; consequently the angular displacement never exceeds $\alpha = \Delta v_y / v^{(0)}_x$. This angular displacement can in principle be measured from the Earth, constituting a detection of a gravitating body in the vicinity of the spacecraft.

%Due to the small size of the typical deflection caused by Planet 9, the angular displacment is maximized rather quickly. 
Treating acceleration due to Planet 9 as instantaneous at $t=0$, the angular displacement varies over time according to
\begin{equation}
\Delta \theta(t) = \left(\frac{t/T}{1 + t/T}\right)\alpha
\text.
\end{equation}
where $T = D / v^{(0)}_x$ is the time it takes for the probe to reach its point of closest approach to the gravitating body. Note that after a total time of $2T$ has passed, the angular displacement is at one-half of its maximum.
The maximum angular displacement attained is given by
\begin{equation}
\alpha \approx 7\times 10^{-9}\;\mathrm{radians}\cdot
\left(
\frac{M}{5 M_\oplus}
\right)
\left(
\frac{40\;\mathrm{AU}}{\rho}
\right)
\left(
\frac{10^{-3}\;c}{v^{(0)}_x}
\right)^2
\text.
\end{equation}
Where this displacement is much larger than the detection threshold, detection occurs before the maximum displacement is reached, and more detailed information about the location of Planet 9 may be extracted from the trajectory.

How large of a displacement is measurable in practice? VLBI is able to measure angular positions at the sub-milliarcsecond level~\cite{lanyi2007angular,1998A&A...329..873K} (corresponding to a detection threshold of order $10^{-9}\;\mathrm{radians}$) for high frequency ($\sim 80\;\mathrm{GHz}$) sources. For lower frequency sources, the angular resolution falls in proportion to the frequency, so it is ideal for the probe to be visible in higher frequencies. Prospects exist for improving this sensitivity by at least an order of magnitude with space-based telescopes~\cite{Fish_2020,An_2020}. Using $10^{-9}$ radians as a detection threshold, we find that gravitating bodies even at the low end of the range of masses for Planet 9 should be detectable at distances of order $100\;\mathrm{AU}$. If a slower probe, with speed $5 \times 10^{-4}\,c$ is used instead of the previously proposed $10^{-3}\,c$, this detection threshold can be quadrupled, at the cost of increasing the mission time to around two decades.

The key manner in which this method is able to increase sensitivity over the original proposal~\cite{witten1} is by use of the fact that the change in transverse velocity is permanent, whereas the change in radial velocity is only nonzero in the vicinity of a gravitating body. The permanent change in transverse velocity allows a large, easily detectable angular displacement to build up over time.

The measurement of angular displacement requires only that the spacecraft be a (perhaps intermittent) source (or reflector) of high-frequency radio waves. Because of the high angular resolution of VLBI, even faint sources can be found by ``resolving out'' the background, reducing the power requirements of transmission. The spacecraft need perform no measurements, nor transmit any information, minimizing the weight and therefore cost of accelerating it to $\sim 0.001\,c$. In particular, this proposal needs no accurate clock on board the probe.

The gravitational influence of Planet 9 is not the only possible source of angular displacement. Other gravitational influences (e.g.\ from planets) and radiation pressures exert forces on the satellite. Radiation pressure from the sun, although large, is along the radial direction, and quite predictable. The other forces at play need to be accounted for, but are not expected to change quickly in the way that the gravitational influence of Planet 9 is. As such, it may not be necessary to have a perfect understanding of these forces before the mission. Furthermore, the satellites that turn out not to pass near Planet 9 can be used instead to map the gravitational potential of the outer parts of the solar system, where there are no visible existing probes.

As discussed in~\cite{witten1}, the Pioneer anomaly~\cite{nieto2003finding} and proposals for testing modified gravity at $\sim 100\;\mathrm{AU}$~\cite{Buscaino_2015} both depend on measuring the radial velocity of a probe. In both cases the primary perturbing force lies along the axis between the observer and the spacecraft, so measuring transverse displacement is not an option. A flyby of Planet 9, however, involves a transverse force as a leading-order perturbation, allowing the use of this different detection method.

\emph{Acknowledgements.} We are grateful to Henry Lamm for comments on an earlier version of this text. S.L.\ is supported by the U.S. Department of Energy under Contract No.~DE-FG02-93ER-40762. Z.R.\ is supported in part by NASA Headquarters under the NASA Earth Science and Space Fellowship grant NNX16AP08H.
\bibliographystyle{unsrt}
\bibliography{p9}
\end{document}